\documentclass[letterpaper,10pt]{article}
\usepackage{graphicx}
\usepackage{amsmath}
\usepackage{amsfonts}
\usepackage{amssymb}
\usepackage{color}
\begin{document}

\title{Magnetic Nanoparticle Assemblies}
\author{Dimitris Kechrakos \footnote{
	Contact author: 
Tel : +30-210-2896705, 
Fax : +30-210-2896713, 
E-mail : \textit{dkehrakos@aspete.gr } } }
\maketitle
\begin{center}
Department of Sciences,
School of Pedagogical and Technological Education (ASPETE),
Athens 14121, Greece
\end{center}
\vspace{0.5in}
\begin{abstract}
This chapter provides an introduction to the fundamental physical ideas and models relevant to the phenomenon of magnetic hysteresis in nanoparticle assemblies. 
The concepts of single-domain particles and superparamagnetism are discussed. The mechanisms of magnetization by coherent rotation and the role of temperature in the gradual decay of magnetization are analyzed in the framework of simple analytical models. Modern numerical techniques (Monte Carlo simulations, Magnetization Dynamics) used to study dense nanoparticle assemblies are presented. An overview of the most common experimental techniques used to measure the magnetic hysteresis effect in nanoparticle assemblies are presented and the underlying principles are exposed.  
\end{abstract}
\vspace{0.5in}
\vspace{1.0in}
\pagebreak
\tableofcontents
\pagebreak

\section{Introduction}

Magnetic nanoparticles (MNPs) are minute parts of magnetic materials with typical size well below $10^{-7}m$. They are present in different materials found in nature such as rocks, living organisms, ceramics and corrosion products, but they are also artificially made and used as the active component of ferrofluids, permanent magnets, soft magnetic materials, biomedical materials and catalysts. Their diverse applications in geology, physics chemistry, biology and medicine renders the study of their properties of great importance both to science and technology.
\par
In geology, the nature and origin of magnetic phenomena related to the presence of magnetite nanoparticles in rocks is of great interest to the palaeomagnetist who searches for the geomagnetic record of rocks.
The presence of magnetite particles associated with the trigeminal nerve in pigeons offers a reliable explanation to the Earth's magnetic field detection and the consequent navigation capability.
In fine arts, magnetic analysis of ancient paintings facilitates the reconstruction of the production techniques of ancient ceramics.
In living organisms, the role of ferritin, a magnetic nanoparticle per se, is important among the iron storage proteins. MNPs are also used as contrast agents in Magnetic Resonance Imaging. Recent work has involved the development of bioconjugated MNPs, which facilitated specific targeting of these MRI probes to brain tumors. MNPs are also used as highly active catalysts which has long been demonstrated by the the use of finely divided metals in several reactions. Owing to their high surface-to-volume ratio MNPs of iron are more efficient at waste remediation than bulk iron.
\par
High density magnetic data storage media provide a major technological driving force for further exploration of MNPs. It is expected that if MNPs with diameter $~5nm$ can be used as individually addressed magnetic bits, magnetic data storage densities of $~1Tbit/in^2$ would be achieved, namely an order of magnitude higher than the present record (Moser 2002). MNPs have also been demonstrated to be functional elements in magneto-optical switches, sensors based on Giant Magneto-Resistance and magnetically controllable Single Electron Transistor devices.
\par
The most common preparation methods for MNPs produce assemblies with different structural and compositional characteristics that depend on the particular method adopted. Granular films, ferrofluids and cluster-assembled films are characterized as assemblies with random order in MNP locations, while ordered arrays are found in patterned media (known also as magnetic dots) and self-assembled films.
The MNP preparation methods are divided to top-down and bottom-up. 
In top-down methods , the NPs are formed from a larger system by appropriate physical processing, such as thermal treatment, etching, etc. 
In bottom-up methods, the NPs are formed by an atomic nucleation process that takes place either in ultrahigh vacuum or in a liquid environment. The latter method relies on colloidal chemistry techniques and presently appears to be the most promising method for production of nanoparticles with extremely narrow size distribution. Colloidal synthesis methods combined with self-assembly methods  produce MNP samples with both size uniformity and long range structural order. It is worth noticing that structural order in a MNP assembly is a decisive property for production of ultrahigh density storage media. 
Owing to their attractive features and their low cost, colloidal synthesis methods  and self-assembly attract presently intense research activity in the field of MNP preparation (Petit 1998, Murray 2001, Willard 2004, Farrell 2005, Darling 2005).
\par
The magnetic properties of MNPs and their assemblies provide a fascinating field for basic research, which is done on two different scales, the atomic and the mesoscopic. In the atomic scale, the properties of individual MNPs are examined and they are revealed in samples with low particle concentration. In the mesoscopic scale, dense samples are examined which exhibit collective magnetic behavior arising from interparticle interactions. 
The study of the magnetic properties can be naturally divided in the investigation of the ground state configuration (long range order, disorder, etc) and the excitations from it. Excitations can be either weak, as for example at low temperature and weak external magnetic field, or strong, as for example, close to a thermal phase transition or under a reversing magnetic field. 

For individual MNPs the ground state configuration can differ remarkably from the parent bulk material in various ways. For example, owing to energy balance reasons, the abundance of magnetic domains that form in a bulk magnet can be replaced by a single domain in a MNP, which then becomes magnetically saturated even in the absence of an external magnetic field (N\'{e}el 1949). The application of an external field forces the atomic magnetic moments of a single-domain MNP to rotate coherently (Stoner 1948). Also, for temperature above a threshold, the direction of particle's magnetization fluctuates at random,  making the particle bahave as a molecule with a giant magnetic moment. The applications of this effect, known as superparamagnetism (Bean 1959), are presently a lot, ranging from geology to medicine. Finally, we should remark that the above described simplified picture of a single-domain MNP becomes invalid if one considers the crucial effect of the MNP surface. Reduced crystal symmetry and chemical disorder close to the surface can produce variations between the surface and interior magnetic structure and modify the overall response of the MNP to an applied field (Kodama 1999). 

When MNPs form dense assemblies, interparticle interactions produce a collective behavior, by coupling the magnetic moments of individual MNPs. This fact renders in most cases even the determination of the ground state configuration an intricate physical problem. The collective behavior of dense (interacting)  assemblies is reflected also on the modified magnetic response of the assembly, compared to isolated MNPs. The most complex behavior occurs in samples with random morphology and long-range magnetostatic interactions. Various experimental measurements have been proposed to reveal the nature of the interparticle interactions, and various measuring protocols probe different aspects of the collective behavior. On the other hand, analytical models have difficulties in predicting or explaining the magnetic behavior of these interacting MNP assemblies, and most of the curret research relies on numerical simulations. 

In this chapter we provide an introduction to the fundamental ideas and concepts pertaining to the magnetic properties of MNP assemblies. Emphasis is given to the response of MNP assemblies to an applied magnetic field and the related issue of magnetization reversal. The chapter is organized as follows : In Section \ref{sec_back} we discuss the magnetic properties of individual (isolated) MNPs. Fist, the condition under which a single-domain MNP is formed is derived, and then the magnetic response under an applied field is examined.  The presentation is based on a simple theoretical model (N\'{e}el 1949, Stoner 1948). In Section \ref{sec_exp_methods} we give a brief overview of the most common magnetic characterization techniques and explain the information extracted from each one. In Section \ref{sec_state} we discuss the response of a dense MNP assembly to a magnetic field, when the interparticle interactions are important and lead to a collective behavior of the MNPs. Mean-field models are presented and an introduction to modern numerical techniques (Monte Carlo, Magnetization dynamics) to tackle this problem are presented. The chapter is summarized in Section \ref{sec_sum} and the perspectives in this field are presented in Sections \ref{sec_persp}. 

\section{Isolated magnetic nanoparticles}\label{sec_back}

In this section we derive the criterion for formation of single-domain MNPs and  examine the magnetization process at zero temperature by coherent rotation of magnetization (Stoner-Wohlfarth model). The behavior of a MNP assembly at finite temperature is discussed and the related concepts of superparamagnetism and blocking temperature are introduced. The effects of an applied dc magnetic field is examined within the simplest model assuming uniaxial anisotropy and bistability of particle moments (N\'{e}el model).

\subsection{Single-Domain Particles}\label{sec_back_SDP}

The ground state magnetic structure of a ferromagnetic (FM) material is the outcome of the balance between three different types of energies, namely, the exchange ($U_{ex}$), the magnetostatic ($U_m$) and the anisotropy energy ($U_a$).
The exchange interaction has its origin in the Pauli exclusion principle for electrons. 
Let the FM material be divided in small cubic elements each one carrying a magnetic moment $\overrightarrow{\mu_i}$. The exchange interaction between the cubic elements favors parallel alignment of neighboring magnetic moments and it is written in the usual Heisenberg form as
$U_{ex}=-(A/a^2)\sum_{ij}cos\theta_{ij}$, where $A$ is the stiffness constant, $a$ is the lattice constant and $\theta_{ij}$ is the angle between moments at  sites $i$ and $j$. The stiffness constant is related to the microscopic exchange energy $J$ through the relation $A=zJS^2/a$, where $S$ is the atomic spin and $z=1,2,4$ for sc, fcc and bcc lattice, respectively.
The magnetostatic energy, is the sum of Coulomb energies between the magnetic moments comprising the FM material. It can be expressed as
$U_m = -\mu_0 \overrightarrow{H_d} \cdot \overrightarrow{M} / 2$,
where $H_d$ is the \textit{demagnetizing} field and $M$ the sample magnetization.
The anisotropy energy, is the energy required to orient the magnetization at an angle ($\theta$) relative to certain fixed axes of the system, known as the \textit{easy} axes. The microscopic mechanisms leading to anisotropy can be quite diverse and the most common types of anisotropy found in FMs are as follows:
\begin{description}
 \item
(i) \textit{Crystal} anisotropy. It arises from the combined effects of spin-orbit coupling and quenching of the orbital momentum that produce a  preferred orientation of the magnetization along a symmetry axes of the underlying crystal.
For a uniaxial materials (e.g. hexagonal Co) it has the form
$U_{a}=K_1 sin^2\theta+K_2 sin^4\theta+ ...$,
where $K_1, K_2,...$ are the anisotropy constants, and $\theta$ the angle between the magnetization direction and the easy axis.
Typical values for cobalt are
$K_1=4.5\times10^6~J/m^3$ and $K_2=1.5\times10^5~J/m^3 $.
For cubic crystals (e.g. fcc Fe, Ni) it reads
$U_a=K_1(a_1^2a_2^2+a_2^2a_3^2+a_3^2a_1^2)+K_2 a_1^2a_2^2a_3^2+...$,
where $a_1,a_2,a_3$ are the direction cosines of the magnetization direction.
Typical values for Fe are $K_1=4.8\times10^4~J/m^3$ and $K_2=\pm0.5\times10^4~J/m^3 $.

\item
(ii) \textit{Stress} anisotropy. It is produced by the presence of stress in the sample and it has a uniaxial character  $U_{a}= K_{\sigma}sin^2\theta$, where $K_{\sigma}=\frac{3}{2}\lambda_i\sigma$, with
$\lambda_i$ the magnetically induced isotropic strain and $\sigma$ the stress.

\item
(iii) \textit{Surface} anisotropy. This is caused by the presence of sample free boundaries, where the reduced symmetry and the presence of defects can induce additional anisotropy. It is important in MNPs because of the substantial surface-to-volume ratio.

\item
(iv) \textit{Shape} anisotropy.
This occurs because on one hand the demagnetizing field depends on the shape of the magnetized body and takes the lowest value along the longest axis of the sample, and on the other hand, $U_m$ is minimized when $M$ is parallel to $H_d$. As an example, consider a specimen in the shape of prolate spheroid with major axis $c$  and minor axis $a$, magnetized at an angle $\theta$ with respect to $c$-axis. Then,
$U_m= \frac{\mu_0}{2}
\left[ N_c(M\cos\theta)^2 + N_a(M\sin\theta)^2 \right] =
\frac{1}{2}(N_c-N_a)M^2\sin^2\theta  $ ,
where $N_c$ and $N_a$ are the \textit{demagnetizing} factors along the corresponding axes. This expression for $U_m$ has the form of uniaxial anisotropy with
$K_s=\frac{1}{2}(N_c-N_a)M^2$. Typical cases, are a spherical specimen with $K_s=0$, an infinitely thin planar specimen with $N_{\Vert}=0$ (in-plane) and $N_{\bot}=1$, and a infinitely long (needle-shaped) specimen with $N_{\Vert}=\frac{1}{2}$ (along the axis) and $N_{\bot}=0$.
\end{description}

In studies of the magnetic properties of MNPs, it is a common practice, to describe, within the simplest approximation, the overall effect of the various anisotropy types by an \textit{effective} uniaxial anisotropy term $U_a=K_{eff}\sin^2\theta$. The constant $K_{eff}$ accounts for the total effect of crystalline, surface and shape anisotropy.

A bulk FM material is composed of many uniformly magnetized regions (\nolinebreak \textit{domains}). The direction of magnetization in different domains varies, and in a bulk sample it is randomly distributed leading to a non-magnetized sample even at temperatures far below the Curie point. 
The formation of magnetic domains in FM materials results from the competition between the exchange and the magnetostatic energy.
The former favors perfect alignment of neighboring moments and the latter is reduced by breaking a uniformly magnetized body into as many as possible regions with opposite magnetization directions.
The outcome of this competition is the formation of a certain number of domains in a sample with a particular orientation of the magnetization directions.
A typical domain size in a bulk ferromagnet is $1\mu m$.

\begin{figure}
\begin{center}
\includegraphics{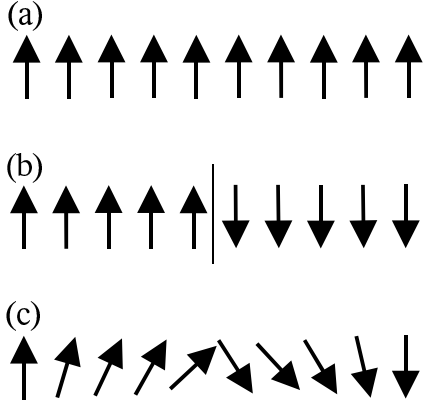}
\caption{One-dimensional model of a FM. (a) Long-range order. (b) An infinitely thin DW (dashed line). The increase of exchange energy at the wall is higher than the decrease of the magnetostatic energy. (c) A $180^0$ domain wall spread over $N=10$ sites. The gradual rotation of atomic moments produces a state with lower total energy compared to (b). }
\label{fig_DW}
\end{center}
\end{figure}

Neighboring magnetic domains are separated by a region where the local magnetization changes gradually direction between the two opposite sides, known as \textit{domain wall} (DW).
Domain walls have finite width $(\delta_w)$ determined by the balance between the exchange and anisotropy energy.
As an example, consider an one-dimensional model of a DW in a uniaxial material, where a $180^0$ rotation of magnetization is distributed over N sites, as shown in Fig.~\ref{fig_DW}.
The total energy per unit area reads
\begin{equation}
\sigma(N)=\sigma_{ex}+\sigma_{a}=JS^2(\pi/N)^2 (N/a^2)+NaK_1.
\label{eq_sN}
\end{equation}
Minimization with respect to $N$ leads to
\begin{equation}
\delta_w=Na=\pi(A/K_1)^{1/2}.
\label{eq_dDW}
\end{equation}
For a typical exchange stiffness value ($A\approx10^{-11}~J/m$), Eq.(\ref{eq_dDW}) predicts for iron
$\delta_w\approx0.4~\mu m$
while for a magnetically harder material like cobalt,
$\delta_w\approx60~nm$.
Substituting the result of Eq.(\ref{eq_dDW}) into in Eq.(\ref{eq_sN}) provides the areal energy density of the DW
\begin{equation}
\sigma_w=2\pi(AK_1)^{1/2}
\label{eq_sDW}
\end{equation}
\par
Consider a finite sample of a FM material, with size $d$. As the size of the sample is reduced, the number of DWs it contains decreases, because fewer regions with opposite directions of magnetization are required to reduce the magnetostatic energy. 
Below a critical value of the system size, the sample does not contain any DW and it is in a \textit{single domain} (SD) state exhibiting saturation magnetization $(M_s)$. 
For a spherical particle, the critical diameter $(d_c)$ can be estimated as follows: the SD state is stable when the energy needed to create a DW that spans the whole particle, $U_w=\sigma_w\pi r^2$, is greater than the magnetostatic energy gain from the reduction to a multidomain state, which is approximately equal to the magnetostatic energy stored in a uniformly magnetized sphere, $U_m=\frac{1}{3}\mu_0M_s^2V$, with $M_s$ the saturation magnetization and $V=\frac{4\pi}{3}r^3$. The condition $U_w=U_m$ provides
\begin{equation}
r_c=9\frac { (AK_1)^{1/2} } {\mu_0M_s^2}
\label{eq_rcr1}
\end{equation}
For Fe, this approximation gives $r_c\approx3~nm$, which is by far too small. The reason is that the DW is assumed to have the same one-dimensional structure as in the bulk material. 
An improved calculation that considers a three-dimensional confinement of the DW  provides for the critical radius:
\begin{equation}
r_c=\sqrt{ \frac{9A}{\mu_0M_s^2}\left[ ln \left( \frac{2r_c}{a} \right) -1 \right]  }
\label{eq_rcr2}
\end{equation}
In the case of Fe, numerical solution of Eq.(\ref{eq_rcr2}) gives $r_c\approx25~nm$, which is very close to more accurate micromagnetic calculations and the experimentally obtained value (Cullity 1972).

\subsection{Magnetization by Coherent Rotation}\label{sec_back_SW}

The magnetization $(M)$ of a bulk FM crystal that contains many magnetic domains, changes under application of an external magnetic field $(H)$, a process known as \textit{technical magnetization}.
However, the value of $M$ is not a unique function of $H$ and the state of the sample \textit{prior} to application of the field is important.
This is the phenomenon of magnetic \textit{hysteresis}, which is commonly depicted by drawing the $M-H$ dependence under a cyclic variation of the field from a positive to a negative and back to a positive saturation value (\textit{hysteresis loop}).
Two important characteristic values of a hysteresis loop are the \textit{remanence} ($M_r$), namely the magnetization after removal of the saturating field, and the \textit{coercivity} ($H_c$), namely the field required for the magnetization to vanish.
In a bulk FM crystal, the magnetization proceeds by two basic mechanisms, namely domain wall motion (weak fields) and rotation of magnetization (strong fields).

\begin{figure}
\begin{center}
\includegraphics{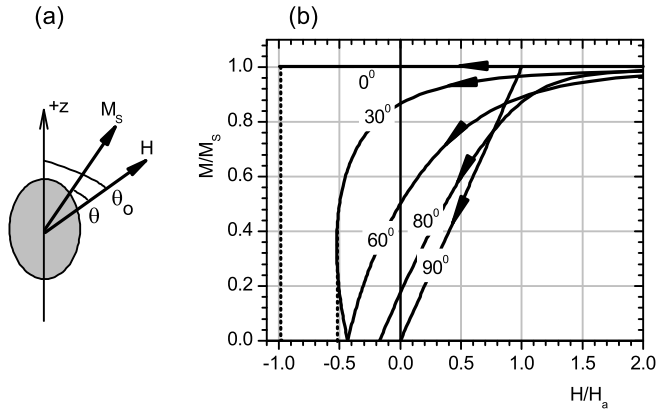}
\caption{
(a) Sketch of a magnetic nanoparticle with uniaxial anisotropy along the $z$-axis and an applied field at an angle ($\theta_0$) with respect to the easy axis.
(b) Magnetization curves within the Stoner-Wohlfarth model for various field directions. The initial direction of the magnetization is taken along the field. }
\label{fig_SW_MH}
\end{center}
\end{figure}

\begin{figure}
\begin{center}
\includegraphics{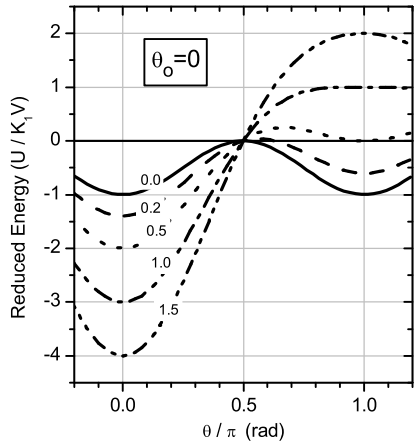}
\caption{
Dependence of total energy on the direction of the particle's moment (see Eq.(\ref{eq_uSD})), for various strengths of the applied field $(h=H/H_a)$. The energy minimum  at $\theta=\pi$ becomes unstable at the switching field $h_s=1$. }
\label{fig_SW-E(theta)}
\end{center}
\end{figure}

In MNPs, the change of magnetization under an applied field proceeds only by rotation, because formation of DWs is energetically unfavorable.
During the magnetization rotation the atomic moments of the MNP remain parallel to each other and the MNP behaves as a giant molecule carrying a magnetic moment of a few thousand Bohr magnetons ($\mu\sim10^4\mu_B$ for a $5~nm$ diameter Fe MNP).
This process of magnetization is known as \textit{coherent} rotation or \textit{Stoner-Wohlfarth} (SW) model, after the authors who introduced and solved it (Stoner 1948). We  discuss it briefly next.
Consider a MNP with uniaxial (effective) anisotropy $K_1$ along an easy axis taken to be the $z$-axis (Fig.~\ref{fig_SW_MH}). For an applied field that makes an angle $\theta_0$ with the easy axis, we wish to determine the equilibrium position of the magnetic moment $\mu=M_sV$. Let $\overrightarrow{\mu}$ make an angle $\theta$ with the easy axis, then the total energy density reads
\begin{equation}
 u=-K_1\cos^2(\theta-\theta_0) - \mu_0 HM_s\cos\theta
\label{eq_uSD}
\end{equation}
The equilibrium condition (zero-torque) is
\begin{equation}
 \dfrac{du}{d\theta}=0  \Rightarrow
2K_1sin(\theta-\theta_0)\cos(\theta-\theta_0) + \mu_0 H M_s \sin\theta = 0
\label{eq_equil}
\end{equation}
and introducing the dimensionless quantity $h=H/H_a$ with the \textit{anisotropy} field 
$H_a=2K_1/\mu_0 M_s$, Eq.(\ref{eq_equil}) becomes
\begin{equation}
 \sin\left( 2(\theta-\theta_0)\right) +2h\sin\theta=0.
\label{eq_equil2}
\end{equation}
We define the reduced magnetization along the field $m=\mu\cos\theta/M_sV=cos\theta$ and the solution of Eq.(\ref{eq_equil2}) is written as
\begin{equation}
 2m(1-m^2)^{1/2}\cos2\theta_0 + \sin2\theta_0(1-2m^2) + 2h(1-m^2)^{1/2}=0
\label{eq_SW_MH}
\end{equation}
The remanence ($h=0$) and coercivity ($m=0$) are readily obtained from Eq.(\ref{eq_SW_MH}) as
\begin{equation}
m_r=\cos\theta_0 \quad \textnormal{and} \quad h_c=\sin\theta_0\cos\theta_0.
 \label{eq_SW_MrHc}
\end{equation}
For non-zero field values, Eq.(\ref{eq_SW_MH}) is solved for $h$ as a function of $m$ and the data are shown in Fig.~\ref{fig_SW_MH}. Consider the two extreme cases, namely for $\theta_0=90^0$ (hard-axis magnetization) and  $\theta_0=0^0$ (easy-axis magnetization). In the former case, the magnetization shows zero coercivity and a linear field dependence. In the latter case, the magnetization remains constant until the reversing field becomes equal to the anisotropy field, and then an \textit{irreversible} jump of the reduced magnetization from $m=+1$ to $m=-1$ is seen. These extreme cases demonstrate the distinct mechanism of switching by rotation that can occur in an assembly. More generally,  at an arbitrary field angle, an irreversible jump of the magnetization occurs at the so called \textit{switching} field ($H_s$) defined as the field value satisfying $dm/dh\rightarrow\infty$. At $H=H_s$ the local minimum of the total energy, corresponding to the higher energy state (magnetization opposite to the applied field) disappears and the system jumps to the remaining minimum that corresponds to a magnetization direction along the field (see Fig.~\ref{fig_SW-E(theta)}). In other words, $H_s$ is an instability point of the total energy and it satisfies $du/d\theta=0$ and $d^2u/d\theta^2=0$. In the SW model, the stability condition reads
\begin{equation}
 \dfrac{d^2 u}{d \theta^2}=0  \Rightarrow
\cos2(\theta-\theta_0) \pm h \sin\theta=0
\label{eq_stab}
\end{equation}
From Eqs.(\ref{eq_equil2}) and (\ref{eq_stab}) we obtain for the switching field $h_s=H_s/H_a$
\begin{equation}
 h_s=(\cos^{2/3}\theta_0 + \sin^{2/3}\theta_0)^{-3/2}
\label{eq_SW_Hs}
\end{equation}
By comparison of Eqs.(\ref{eq_SW_MrHc}) and (\ref{eq_SW_Hs}) one finds that $h_c < h_s$ for $45^0 < \theta_0 < 90^0$ , namely switching happens after the magnetization changes sign, while for field angles close to the easy axis, $0^0 < \theta_0 < 45^0$, the magnetization changes sign by an irreversible jump ($h_c=h_s$).
The physical distinction between $h_s$ and $h_c$ can be understood by the following example.
Consider a SW particle under application of a reversing field $h=h_c$, which brings the particle's moment $\overrightarrow{\mu}$ in a direction perpendicular to the field, so that $m=0$.
Then the field is switched off adiabatically. If $h_c<h_s$ (i.e. $45^0 < \theta_0 < 90^0$), $\overrightarrow{\mu}$  will return back to the positive remanence value $(m=+1)$, while if $h_c=h_s$ (i.e. $0^0 < \theta_0 < 45^0$), $\overrightarrow{\mu}$  will jump to the negative remanence state $(m=-1)$.
The switching field of a hard (i.e. large anisotropy) magnetic material is a physical quantity with great technological interest in magnetic recording applications. In these, the information bit is stored in the direction of magnetization and the switching field is the field required to write or erase this information.

Stoner and Wohlfarth (Stoner 1948) also studied an assembly of isolated MNPs with  easy axes directions distributed uniformly on a sphere (\textit{random anisotropy model}, RIM). The reported values for the remanence and coercivity are
\begin{equation}
m_r=0.5 \quad \textnormal{and} \quad h_c=0.48.
 \label{eq_RAM}
\end{equation}
This result is particularly useful as random easy axis distribution is found in most MNP-based materials (granular films, cluster-assembled films, self-assembled arrays, etc)

As a final remark, we remind that in the SW model thermal effects are ignored $(T=0)$, thus energy-minimization with respect to the magnetic moment direction is a sufficient condition to determine the field-dependent magnetization at equilibrium. The magnetic behavior of SD particles at finite temperature is discussed in the following section.

\subsection{Magnetic behavior at finite temperature}

How do thermal fluctuations affect the average magnetization direction of an isolated MNP~?
How does the presence of an applied field modify the magnetic response at finite temperature~?
Is the assembly magnetization stable in time, when the MNP moment are subject to thermal fluctuations~?
These points are briefly discussed next, along the lines of a model first studied by N\'{e}el (N\'{e}el 1949).

\subsubsection{Superparamagnetism and Blocking temperature}
\label{sec_back_SPM}

Consider an assembly of identical SD particles with uniaxial anisotropy.
The energy (per particle) is $U=-K_1V\cos^2\theta$, where $\theta$ is the angle between the single particle magnetic moment $\overrightarrow{\mu}$ and the easy axis. The energy barrier that must be overcome for a MNP to rotate its magnetization is $E_b=K_1V$.
As first pointed out by N\'{e}el (N\'{e}el 1949), thermal fluctuations could provide the required energy to overcome the anisotropy barrier and spontaneously (i.e. without externally applied field) reverse the magnetization of a MNP from one easy direction to the other.
This phenomenon can be thought of as a Brownian motion of a particle's magnetic moment.
The assembly shows paramagnetic behavior, however it is the giant moments of the MNPs that fluctuate rather than the atomic moments of a classical bulk paramagnetic material.
This magnetic behavior of the MNPs is called \textit{superparamagnetism} (SPM) (Bean 1959)
At high enough temperature, $k_BT>>K_1V$, the anisotropy energy can be neglected and the assembly  magnetization can be described by the well known Langevin function $M =nM_s \L(x)$, where $n$ is the particle number density, and $x=\mu_0\mu H/k_BT$.
Thus, the features serving as signature of superparamagnetism are the scaling of magnetization curves with $H/T$, as dictated by the Langevin function, and the lack of hysteresis, i.e. vanishing remanence and coercivity.
Moreover, the major difference between classical paramagnetism of bulk materials and SPM is the weak fields $(H\sim0.1~T)$ required to achieve saturation of a MNP assembly magnetization $M$.
This occurs because of the large particle moment $(\mu \sim 10^4 \mu_B)$ compared to the atomic moments $(\mu_{at} \sim \mu_B)$.

Measurement of magnetization curves at sufficiently high temperature can, in principle,  be used to extract the particle moment $\mu$. In practice, two complication arise. First, the presence of different particle sizes in any sample produces a convolution of the Langevin function with the volume distribution function. Second, interparticle interactions, modify the reversal mechanism and the SW model needs extensions, which are discussed in the Section~\ref{sec_state}.

At low temperature, $k_BT<<K_1V$, the anisotropy barriers are very rarely overcome (weak thermal fluctuations), the assembly shows hysteresis and this is called the \textit{blocked} state.

One might now ask, whether there exists a temperature value that draws the border between the blocked and the SPM state. Following N\'{e}el's arguments, we assume that thermal activation over the anisotropy barrier can be described within the \textit{relaxation time} approximation (or \textit{Arrhenius law}) as
\begin{equation}
\tau=\tau_0\exp(K_1V/k_BT),
\label{eq_tau}
\end{equation}
where $1/2\tau$ is the probability per unit time for a reversal of $\overrightarrow{\mu}$. The intrinsic time $\tau_0$ depends on the material parameters (magnetostriction constant, Young modulus, anisotropy constant and saturation magnetization). Typical values are $\tau_0\sim 10^{-10}-10^{-9}s$ as obtained by N\'{e}el.
To detect the superparamagnetic behavior experimentally, the MNP must be probed for a long enough period of time to perform many switching events that would produce a vanishing small time-average magnetic moment.
If $\tau_m$ is the measuring time-window, the condition for SPM behavior is $\tau_m \gg \tau$.
The strong (exponential) dependence of $\tau$ on temperature (see Eq.(\ref{eq_tau})) permits us to define a temperature value (or more precisely, a very narrow temperature range) above which the relaxation time is so small that SPM behavior is observed.
This is called the \textit{blocking temperature} ($T_b$) of the assembly, and is given by
\begin{equation}
 T_b=K_1V / k_B \ln(\tau_m / \tau) .
\label{eq_Tb}
\end{equation}
For $T<T_b$, the particle moments fluctuate without switching direction (on average) and the assembly is in the blocked state exhibiting hysteresis.
For $T>T_b$ the assembly is in the SPM state, hysteresis disappears and thermal equilibrium is established.
It is remarkable, that the value of $T_b$ depends on $\tau_m$, which is a characteristic of the experimental technique adopted.
For example, in dc susceptibility measurements
$\tau_m\approx 100s$,
in ac susceptibility
$\tau_m \approx 10^{-8}-10^4s$,
in M\"{o}ssbauer spectroscopy
$\tau_m \approx 10^{-9}-10^{-7}s$
and in neutron spectroscopy
$\tau_m \approx 10^{-12}-10^{-8}s$.
Therefore, if $T_b$ is of interest for a particular application, the measurement technique implemented must imitate the real conditions. For example, to study the reliability of magnetic storage media, dc magnetic measurements over a wide time window ($\tau_m\sim 10^2-10^4~s$) should be used, while to study magnetic recording speed, ac measurements are appropriate.

Brown (Brown 1963) extended the treatment of thermal activation over the anisotropy barrier, allowing also for fluctuations of $\mu$ transverse to the easy axis, which N\'{e}el has neglected, and obtained a different expression for $\tau_0$.
However, the common feature of both studies is the temperature and volume dependence of $\tau$, so the final result, Eq.(\ref{eq_tau}), is referred to as the \textit{N\'{e}el-Brown} model.

In a \textit{polydisperse} assembly, the distribution of particle volumes $f(V)$,  produces a corresponding distribution of blocking temperatures $f(T_b)$. Then, at a certain temperature $T$ the assembly contains a mixture of blocked and SPM particles.
The MNPs with volumes above a critical value $V_c$, fulfill the requirement of strong thermal energy with respect to their anisotropy barrier, and are SPM, while those while those with $V\leq V_c$ are blocked.
From Eq.(\ref{eq_tau}), the critical volume reads $V_c=k_bT\ln(\tau_m/\tau_0)/K_1$. 
As explained above for $T_b$, also for $V_c$ the experimental determination  depends on the technique adopted.
Most preparation techniques result in polydisperse samples and the problem of extracting the size distribution function from magnetic measurements, pioneered by Bean and Jacobs more than fifty years ago (Bean 1956) remains a difficult task mainly due to the complications introduced by interparticle interactions.
Knobel and colleagues have recently reviewed this subject (Knobel 2008).

\subsubsection{Thermal relaxation under an applied field}
\label{sec_back_relax_H}

\begin{figure}
\begin{center}
\includegraphics{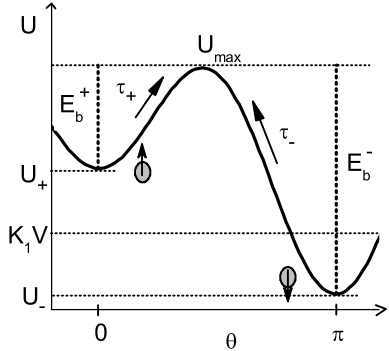}
\caption{Total energy of an isolated particle with uniaxial anisotropy subject to a negative field parallel to the easy axis with value less than the switching field $(0<H<H_s)$. Energy barriers $(E_b)$ and relaxation times $(\tau)$ for the forward $(+)$ and the backward $(-)$ process are not equal.}
\label{fig_SW_tau}
\end{center}
\end{figure}
\par
Consider an assembly of $N$ identical MNPs with uniaxial anisotropy along the $z$-axis and let their moments point initially along the $+z$-axis.
Assume that  a magnetic field $H$, weaker than the switching field , which is equal to $H_a$, is applied along the $-z$-axis.
Then, the total energy per particle reads,
$U=-K_1V\cos^2(\theta-\theta_0) + \mu_0 HM_sV\cos\theta$.
It exhibits two non-equivalent local minima at $\theta=0, \pi$ with values
$U_{\pm}=-K_1V \pm M_sVH$ and a maximum at $\theta=\pi/2$ with $U_{max}=K_1V(H/H_a)^2$, as shown in Fig.~\ref{fig_SW_tau}.
The energy barriers and the corresponding relaxation times for the forward $(+)$
and the backward $(-)$
rotations are
\begin{equation}
E_b^{\pm}(H)=K_1V(1 \mp H/H_a)^2
\quad \textnormal{and} \quad
\tau_{\pm}=\tau_0\exp({E_b^{\pm}/k_BT}).
\label{eq_tau(H)}
\end{equation}
The change of $\tau_0$ due to the field is much weaker than the change of the exponential factor and as such it is neglected in the above equation.
\par
The blocking temperature, as measured within a time-window $\tau_m$, is reached when the observation time equals the \textit{forward} relaxation time $\tau_+$, because the latter corresponds to a moment flip from the initial state along $+z$ to the opposite direction, namely a process that reduces the initial magnetization.
From Eq.(\ref{eq_tau(H)}) one obtains
\begin{equation}
 T_b(H)=\frac{K_1V(1-H/H_a)^2}{k_B \ln(\tau_m / \tau)}
       \equiv T_b(0) ( 1-H/H_a ) ^2.
 \label{eq_Tb(H)}
\end{equation}
which indicates that the blocking temperature is reduced by the presence of a reverse field.
By completely symmetric arguments one could show that $T_b$ increases in the presence of a field with the same direction as the initial magnetization.
\par
Since thermal fluctuations act in synergy to a reverse field in switching the moment of a MNP, it is expected that the coercivity of an assembly will decay with temperature.
As discussed above, for a particle with its moment along the $+z$-axis, a reverse field $(0 < H < H_a)$ reduces the barrier for reversal to the value $E_b^+$ given in Eq.(\ref{eq_tau(H)}).
If the field is strong enough, it will reduce the barrier to the value appropriate for superparamagnetic relaxation, namely $k_BT \ln (\tau_m/\tau_0)$, and the (time-average) magnetization will vanish. On the other hand, the reverse field that makes the magnetization vanish is by definition the coercive field. Therefore, the following relation holds
\begin{equation}
K_1V(1 - H_c/H_a)^2 = k_BT \ln (\tau_m/\tau_0)
\end{equation}
which, using Eq.(\ref{eq_Tb}), provides the temperature dependent coercivity
\begin{equation}
 H_c(T)=H_a \left[ 1-(T/T_b)^{1/2} \right].
\label{eq_SW_Hc(T)}
\end{equation}
\par
The microscopic mechanism of thermal activation of the MNP moment over the anisotropy barrier, produces a  macroscopically measured time-decay of the magnetization.
We derive this dependence assuming that when a moment switches direction it continues to remain along the easy axis  (N\'{e}el 1949).
Then, at time $t$, $N_+$ particles occupy the lower minimum at $\theta=0$, and the rest $N_-=N-N_+$ particles occupy the higher minimum at $\theta=\pi$.
The time-evolution of $N_+$ is governed by the \textit{rate equation}
\begin{equation}
 \dfrac{dN_+}{dt}=-\dfrac{N_+}{\tau_+}+\dfrac{N_-}{\tau_-}.
 \label{eq_rate_N+(H)}
\end{equation}
The magnetization per particle is given as
$M(t)\equiv(2N_+(t)/N - 1)M_s$,
and solution of Eq.(\ref{eq_rate_N+(H)}) provides
\begin{equation}
 M(t)=M_{\infty} + (M_0-M_{\infty})\exp(-t/\tau)
 \label{eq_M(t,H)}
\end{equation}
with $1/\tau=1/\tau_+ + 1/\tau_-$
being the reduced relaxation time and
\begin{equation}
 M_{\infty}=\dfrac{\tau_+ - \tau}{\tau_+ + \tau_-}M_s
\quad \textnormal{and} \quad
M_0=\left( \dfrac{2N_+(0)}{N_-(0)}-1  \right) M_s
\end{equation}
the time-asymptote and initial values of the particle magnetization, respectively.
Eq.(\ref{eq_M(t,H)}) indicates that the magnetization decays exponentially towards the equilibrium value $M_{\infty}$, reached as $t\rightarrow\infty$.
In other words, equilibrium is reached when the population of the energy minima is proportional to the corresponding relaxation times
$(N_+/N_-=\tau_+/\tau_-)$,
as dictated by Eq.(\ref{eq_rate_N+(H)}).
When the applied field is strong enough $(H>H_s)$ to produce only one minimum, thermal equilibrium is always reached.
Obviously, in the absence of an external field, thermal equilibrium is reached when the two equivalent minima are equally populated $(N_+=N_-)$, resulting in a vanishing magnetization.

Notice that in Eq.(\ref{eq_rate_N+(H)}) we assumed bistability of the moment direction, which is a valid approximation provided the anisotropy barrier is high $(K_1V\gg k_BT)$. For lower anisotropy barriers or elevated temperature $(K_1V \approx k_BT)$, the transverse fluctuations of $\overrightarrow{\mu}$, or, in other words, intra-valley motion around the energy minimum should be taken into account. A general treatment of thermal relaxation of SD MNPs was pioneered by Brown (Brown 1963) and extended to the case of an applied external field (Aharoni 1965, Coffey 1998, Garannin 1999).
\par
If an assembly is polydisperse, characterized by a volume distribution $f(V)$, a distribution of blocking temperatures $f(T_b)$ exists. However, it remains still unclear if the mean value $<T_b>$ is the appropriate blocking temperature of the assembly, which should be substituted, for example, in Eq.(\ref{eq_SW_Hc(T)}). This point is discussed further in the literature (Nunes \textit{et al} 2004).

In a polydisperse assembly, a distribution of relaxation times $f(\tau)$ exists, with $f(\tau) d(\ln\tau) $ the probability of a MNP to have $\ln\tau$ in the range
$\left( \ln\tau, ~\ln\tau+d\ln\tau \right)$ and the normalization condition
$\int_0^{\infty} f(\tau) d(\ln\tau) =1$. In this case the magnetization can be obtained by a  superposition of the single-particle magnetization properly weighted, as follows
\begin{equation}
 M(t)=M_s \int_0^{\infty} \left[ 1-\exp(-t/\tau)  \right]
                           \frac{f(\tau)}{\tau} d\tau,
\end{equation}
where the term in brackets is the probability per unit time for a particle not to flip its moment.
For a broad enough distribution,  the observation time $t$ will satisfy $\tau_1\ll t \ll \tau_2$, where $\tau_1$ and $\tau_2$ are the minimum and maximum relaxation times of the assembly, respectively.
Assuming a uniform distribution $f(\tau)$, it can be shown that the magnetization exhibits a logarithmic relaxation
\begin{equation}
M(H,t)=M(H,0)- S(H,T) \ln (t/\tau_0)
\label{eq_M2(t,H=0)}
\end{equation}
with $S$ the \textit{magnetic viscosity} of the system. Thus, polydispersity produces a much slower decay of magnetization with time.
\par
The discussion so far, refers to a field applied parallel to the easy axis. However, random anisotropy is most commonly found in MNP assemblies and the the necessity to study the effect of a tilted field with respect to the easy axis, arises.
In this case, the calculation of the energy barriers and relaxation time is a much more complicated task and no analytical solution exists. Numerical studies (Pfeiffer 1990) showed that the energy barrier for an applied field at an angle $\theta_0$ to the easy axis can be approximately written as 
\begin{equation}
 E_b(\theta_0)=K_1V ( 1 - H/H_a)^{0.86+1.14h_s}
 \label{eq_Pfeif_Eb}
\end{equation}
where $h_s$ is given by Eq.(\ref{eq_SW_Hs}). In the limit of $\theta_0=0$, Eq.(\ref{eq_Pfeif_Eb}) reduces to Eq.(\ref{eq_tau(H)}).

The temperature dependence of the coercivity for a monodisperse assembly with random anisotropy has also been obtained numerically (Pfeiffer 1990) as
\begin{equation}
 H_c(T)=0.48H_a \left[ 1-(T/T_b)^{0.77} \right],
\label{eq_RAM_Hc(T)}
\end{equation}
which at $T=0$ reduces to the SW result of Eq.(\ref{eq_RAM}).
A detailed theoretical study of the relaxation time for a non-uniaxial applied field can be found in the review by Coffey and colleagues (Coffey 1993).
\par
As a concluding remark, the presence of polydispersity and random anisotropy makes the description of the magnetic behavior of an assembly intractable to exact analytical treatment. Instead, numerical approximations and simulation methods provide the alternative theoretical tools to study these systems. Numerical simulation approaches are introduced in Section \ref{sec_state}.

\section{Magnetic Measurements} \label{sec_exp_methods}

Thermal relaxation has a dynamic character, therefore, the relation between the various relaxation times of the assembly and the measurement time is a decisive parameter for the outcome of a measurement.  Additionally, if the assembly is not at equilibrium during the measurement, or if it changes its equilibrium state (for example, by adiabatic changes of the applied field) the result of the measurements depends on the measurement protocol followed.
In what follows we discuss two very common types of static measurements, that reveal the temperature and field dependence of the magnetization and provide evidence for superparamagnetic relaxation.
Dynamic measurements are not discussed in this article. The interested reader can find more on the physical principles behind the most common magnetic measurement techniques in the review of Dormann and colleagues (Dormann 1997).
\par
\subsection{Field-cooled (FC) and Zero-Field-Cooled (ZFC) Magnetization}
\begin{figure}
\begin{center}
\includegraphics{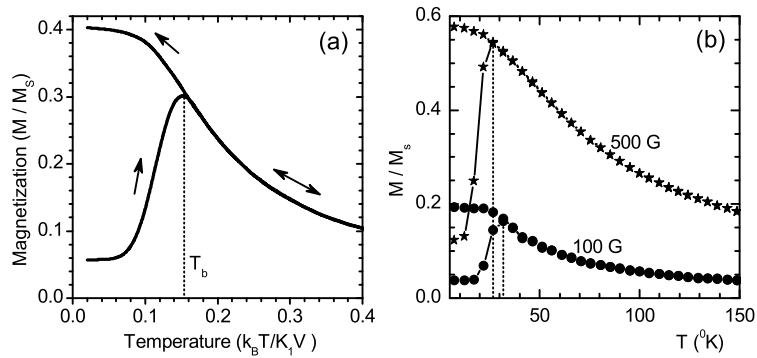}
\caption{(a) Typical FC-ZFC magnetization curves. The curves join at the peak of the ZFC which corresponds to $T_b$. Arrows indicate the direction the measurements are taken. For $T>T_b$ the system is in thermal equilibrium and the heating process is reversible. (b) FC-ZFC curves for a dilute (non-interacting) assembly of Fe nanoparticles ($D=3.0~nm, M_s=1720~emu/cc$ and $K_1=2.4\times 10^5~erg/cc$). The blocking temperature (dotted line) decreases weakly with increasing measuring field. The non-zero values of $M_{ZFC}(T=0)$  are due to the finite value of the measuring field. Data produced by Monte Carlo simulations (Section \ref{ssec_MCMD})  }
\label{fig_ZFC}
\end{center}
\end{figure}

This is a  measurement protocol adopted for investigation of the temperature dependent magnetization of an assembly and it reveals superparamagnetic behavior.
It is performed in three stages. 
In the first, the sample is initially at a high enough temperature ($T_{max}$) to ensure a SPM state and it is cooled to low temperature ($T_{min}$) to approach its ground state.
In the second stage, a weak field is applied ($H\ll H_{sat}$), the sample is heated up to $T_{max}$ and the magnetization is measured as a function of temperature.
This is the ZFC curve.
In the third stage, the system is cooled down to $T_{min}$, without removing the field, while the magnetization is recorded again, producing the FC curve.
During cooling and heating the temperature changes at the same constant rate. 
A typical ZFC-FC curve is shown in Fig.~\ref{fig_ZFC}.
As the temperature rises, the blocked magnetic moments align easier along the applied field leading to an initial increase of the ZFC curve.
However, as soon as thermal fluctuations push the moments over the anisotropy barrier, thermal randomization of the moments produces a drop of the curve.
Therefore, the peak of ZFC curve corresponds to the blocking temperature of the assembly.
Notice that above $T_b$ the ZFC and FC curves coincide, because the system is in thermal equilibrium and the the heating (cooling) process is reversible.
On cooling below $T_b$ the moments remain partially aligned along the field, and the magnetization tends to a non-zero value.
The magnetization vanishes at the ground state ($T_{min}$) if the measuring field is very weak, a random distribution of the easy axes exists and the assembly is non-interacting (dilute).
Deviations from any of the above conditions produce a non-zero value for $M_{ZFC}(T=0)$.
For isolated MNPs, the ZFC-FC curves are only weakly sensitive to the value of the applied field, provided that it is weak $(H \ll H_s)$.
\par
\subsection{Remanent magnetization and Coercive field}
Remanent magnetization at a certain field, $M_r(H)$, is measured after switching off the previously applied field $H$. 
In an assembly of MNPs, remanence arises because the moments of some particles, which have rotated under an applied field and to do so they have overcome an energy barrier, cannot rotate back to their original direction after removal of the field.
In a polydisperse assembly, at finite temperature $T$, only the blocked MNPs, namely those with $T_b < T$ contribute to the remanence.
Therefore,
 $M_r/M_s=\int_{E_{b,c}}^\infty f(E_b) dE_b$
where $E_{b,c}=K_1V_c$ is the critical barrier for SPM relaxation at temperature $T$.
Taking into account that  $T_b\sim V$ (see Eq.(\ref{eq_Tb})), we deduce that
\begin{equation}
 dM_r(T)/dT = f(T_b)
\end{equation}
namely, the slope of $M_r(T)$ provides the barrier (or blocking temperature) distribution function of the assembly.
There are three different measurement protocols for the remanent magnetization, as first suggested by Wohlfarth (Wohlfarth 1958) :
\begin{description}
\item
(i) \textit{Thermoremanence} $TRM(H,T)$, measured at the end of a FC process with field $H$ from $T_{max}$ down to the measuring temperature $T$.
\item
(ii) \textit{Isothermal Remanence} $IRM(H,T)$, measured at the end of ZFC process from $T_{max}$ down to the measuring temperature $T$, at which a field $H$ is applied and then removed.
\item
(iii) \textit{DC Demagnetization remanence} $DcD(H,T)$. First, a ZFC process from $T_{max}$ down to the measuring temperature $T$ is performed. Second, the sample is brought to \textit{saturation} remanence $IRM(\infty,T)$. Third, a \textit{reverse} field $H$ is applied and then removed to leave the sample at the $DcD(H,T)$ remanence.
\end{description}

Wohlfarth pointed out that for isolated MNPs the different remanent magnetizations are related as $DcD(H) = IRM(\infty) - 2\cdot IRM(H)$.
More interestingly, the deviations from this equality, defined as
\begin{equation}
 \Delta M(H) = DcD(H) - \left[ IRM(\infty) - 2\cdot IRM(H) \right]
\label{eq_Henkel}
\end{equation}
quantify the character  and strength of interparticle interactions and are obtained experimentally (O'Grady \textit{et al} 1993).
Positive $\Delta M$ values imply interactions with magnetizing character, and negative values indicate demagnetizing interactions.
We should say that this is only a phenomenological characterization of the interactions, because Eq.(\ref{eq_Henkel}) does not provide any information about their microscopic origin.
However, Eq.(\ref{eq_Henkel}) has been proved a standard tool for quantification of interparticle interactions in complex MNP assemblies such as those used in modern industry of magnetic recording media (granular films, particulate media). 
Interparticle interactions are discussed in the  Section~\ref{sec_state}.

\section{Interacting nanoparticle assemblies}  \label{sec_state}

\subsection{Introduction}
The magnetic interactions that are present in bulk magnetic materials pertain to MNP assemblies and they preserve their physical origin and characteristics.
In particular, (direct) exchange between atomic moments separated by a few lattice constants can couple ferromagnetically or antiferromagnetically two MNPs via their surface atoms.
Indirect exchange or Ruderman-Kittel-Kasuya-Yosida (RKKY) interaction exists between MNPs hosted in a metallic matrix, which provides free electrons required to mediate the interaction between the atomic moments of the MNPs.
Finally, magnetostatic interactions, which are of minor importance in bulk magnets due to their weakness, become the dominant interactions in MNP assemblies with \textit{well separated} MNPs.
This situation occurs for two reasons.
First, the exchange interactions have a very short range (up to $\sim 5~$\AA), the RKKY interactions have an oscillating FM/AFM character with a period of a few \AA, which renders to zero their average effect on the MNP volume, so both have a weak effect in interparticle coupling.
On the other hand, magnetostatic interactions, in the lowest approximation, namely the dipolar contribution, are proportional to the magnitude of the coupled magnetic moments, which for SD particles has an enormously large value compared to the atomic moments
($\mu_{MNP}\sim 10^3\mu_B \sim 10^3\mu_{atom} $).
\par
\begin{figure}
\begin{center}
\includegraphics{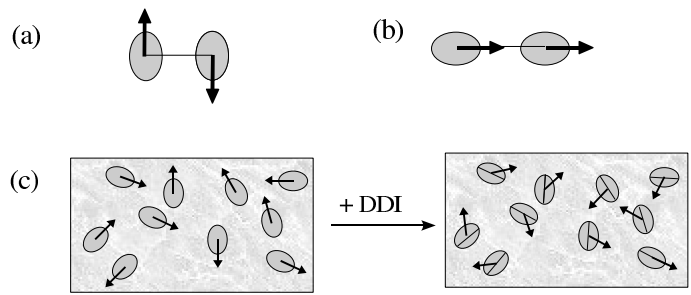}
\caption{Ground state configuration of magnetic nanoparticles with elliptic shape coupled by magnetostatic (dipolar) forces. The easy axis coincides with the long axis of the ellipse (shape anisotropy). Dipolar coupling induces antiferromagnetic ordering when the moments are forced (by anisotropy) to remain normal to the bond, as in (a). FM ordering (nose-to-tail) is favored when the easy axes are parallel to the bond, as in (b). In an assembly with random anisotropy, as in (c), the moments are aligned along the local easy axes. Dipolar interactions have a complex effect, leading to misalignment of the moments with respect to the local easy axis.}
\label{fig_DDI}
\end{center}
\end{figure}

Further on we discuss the effects of magnetostatic (dipolar) interactions on the magnetic properties of MNP assemblies and their interplay with single-particle anisotropy.
The complexity of this problem arises from the \textit{long-range} ($\sim1/d^3$, $d$ being the interparticle distance) and \textit{anisotropic} character of the dipolar interactions, namely the dependence of interaction energy on the orientation of the moments relative to the bond joining the particle centers (Fig.~\ref{fig_DDI}).
\par
Understanding and controlling the effects of dipole-dipole interactions (DDI) in MNP assemblies is of paramount importance to modern technology of magnetic recording media for two opposite reasons. First, DDI couple the MNPs of an assembly. 
The ultimate goal in magnetic recording applications is to address each MNP individually and treat it as a magnetic bit.
In this case, DDI have a parasitic role and one wishes to estimate and reduce their impact in the magnetic properties of an assembly. 
On the contrary, magnetic logic devices, have been proposed and built that exploit the magnetostatic coupling between ordered MNP arrays (linear or planar) to transfer a magnetic bit (usually a flipped moment) between two distant points in the array (Cowburn 2006). 
In this case, DDI are of central importance and the goal is to enhance and tailor their effects.
\par
Over the last two decades, many research groups have prepared and measured MNP assemblies in various forms (granular films, ferrofluids, cluster assembled films,  self-assembled nanoparticles, lithographic arrays of magnetic dots) and studied the intrinsic factors (host and particle material, particle size, particle density) and the extrinsic factors (temperature, field, measurement protocol) that control the magnetic behavior. In many of these studies the presence of magnetostatic interactions has been confirmed.
Among the above mentioned systems, the self-assembled MNPs prepared by a synthetic route offer the advantage of containing well separated MNPs with a very narrow size distribution ($\sigma_V\sim5-10\%$), so they are ideal systems to study DDI effects.
Experimental observations on self-assembled MNPs that have been  attributed to DDI include,
reduction of the remanence at low temperature (Held 2001),
increase of the blocking temperature (Murray 2001),
increase of the barrier distribution width (Woods 2001),
deviations of the zero-field cooled magnetization curves from the Curie behavior (Puntes 2001),
and difference between the in-plane and out-of-plane remanence (Russier 2000).
Long-range ferromagnetic order in linear chains (Russier 2003),
and hexagonal arrays (Puntes 2004, Yamamoto 2008)
of dipolar coupled single-domain magnetic nanoparticles
has been demonstrated, supporting the existence of a \emph{dipolar superferromagnetic} ground state, characterized by ferromagnetic long-range order of the particle moments.

Investigations of the static and dynamic magnetic properties of dipolar interacting nanoparticle assemblies brought up fundamental issues related to the existence of a ground state which shares common features with \textit{spin glasses}, such as slow relaxation, memory and ageing effects (Sasaki 2005).
The latter are magnetic systems characterized by disorder and competing interactions that produce an energy landscape with many local minima, considered responsible for the occurrence of these effects. Dipolar interparticle interactions in dense and random nanoparticle assemblies are believed to cause a \textit{spin-glass-like} behavior (Dormann 1997).

Theoretical models have been developed in an effort to explain these observations and related previous ones in assemblies with randomly located MNPs (granular films, cluster-assembled films).
On a microscopic level, the presence of DDI between MNPs modifies the  magnetization switching mechanism, which for an isolated MNP obeys the N\'{e}el-Arrhenius model.
When anisotropic MNPs are dipolar coupled, the reversal mechanism is determined by the interplay between the single-particle anisotropy energy ($E_a \sim K_1V$) and the dipolar interaction energy ($E_d\sim \mu_i\mu_j/r_{ij}^3$).
For weak interactions ($E_d \ll E_a$), the moments reverse \textit{independently} by thermal activation over energy barriers, which are however modified due to DDI.
This limiting case is treated within a mean-field approximation and  is discussed in Section~\ref{ssec_MFT}.
For strong interactions ($E_d \gg  E_a$), the single-particle reversal is no longer valid.
Reversal of one particle can excite the reversal of others, and the assembly behaves in a collective manner.
Many-body energy barriers exist in the system, with values that depend on the configuration of all moments.
Their evaluation becomes a formidable task and numerical simulations offer in this case an indispensable tool.
Numerical methods are briefly discussed in Section \ref{ssec_MCMD}.

For a detailed review on the magnetic properties of dipolar interacting MNP assemblies the reader is referred to the relevant literature (Dormann 1997, Farrell \textit{et al} 2005, Knobel \textit{et al} 2008, Kechrakos and Trohidou 2008). The role of magnetostatic interactions in patterned magnetic media has been reviewed by Mart\'{\i}n \textit{et al} (Mart\'{\i}n 2003)

\subsection{Mean Field models}\label{ssec_MFT}

In an early attempt to include the effect of interparticle interactions in the thermal relaxation of MNPs , Shtrikman and Wohlfarth (Shtrikman 1981) assumed that the single-particle anisotropy barrier of a MNP is increased by the Zeeman energy due to the \textit{interaction field} $H_{int}$ produced by the moments of neighboring particles.
In this model, the N\'{e}el relaxation time is obtained from Eq.(\ref{eq_tau(H)}) with the applied field $H$ replaced by the interaction field $H_{int}$.
The mean-field approximations consists in replacing $H_{int}$ by its thermal average value, which in  N\'{e}el's model is
\begin{equation}
 \overline{H_{int}} = H_{int} \tanh(\mu_0\mu H_{int}/k_BT)\approx 
 \mu_0\mu \overline{H_{int}^2}/k_BT
\label{eq_Hi}
\end{equation}
the latter approximation being valid for weak interaction fields. Substitution of Eq.(\ref{eq_Hi}) into Eq.(\ref{eq_tau(H)}) gives
\begin{equation}
 \tau
=\tau_0\exp \left[ \frac{K_1V  + \mu_0^2\mu^2 \overline{H_{int}^2}/k_BT }{k_BT} \right].
\label{eq_tau_DDI}
\end{equation}
Using the approximation $1+x \approx 1/(1-x)$ we write Eq.(\ref{eq_tau_DDI}) in the form
\begin{equation}
 \tau
\approx
\tau_0\exp \left[ \frac{K_1V}{k_B(T-T_0)}  \right]
\label{eq_Vogel}
\end{equation}
with $T_0=\mu_0^2\mu^2 \overline{H_{int}^2} / k_BK_1V$.
Eq.(\ref{eq_Vogel}), also known as the \emph{Vogel-Fulcher law}, indicates that the relaxation time of an assembly of interacting MNPs is the same as that of the isolated MNPs at a \textit{lower} temperature.

In the Shtrikman-Wohlfarth model, the temperature $T_0$, or equivalently the thermal average $\overline{H_{int}^2}$, is not related to the microscopic parameters of the assembly, i.e. particle location, and is treated as a phenomenological parameter, fitted to experimental data.
Dormann and colleagues (Dormann 1988, Dormann 1997) developed a statistical model for the average barrier in a dipolar interacting assembly which quantifies the interaction field and provides for the single-particle energy barrier
\begin{equation}
 E_b=K_1V + n_1 a_1 M_s^2 V
\L (a_1 \mu^2 / V k_BT)
\label{eq_DFB_model}
\end{equation}
with $n_1$ the number of nearest neighbors of a particle,  $a_1=x_v/ \sqrt{2}$, $x_v$ the volume concentration of the particles and $\L (\cdot)$ the Langevin function.
Eq.(\ref{eq_DFB_model}) indicates that the anisotropy barrier is increased due to DDI, thus the model of Dormann \textit{et al} predicts an increase of the blocking temperature due to DDI. This model behavior has been observed in almost all types of MNP assemblies, with a few exceptions (Hansen and M{\o}rup 1998).

More recently, Allia \textit{et al} (Allia 2001) used also a phenomenological approach to describe a superparamagnetic assembly with weak DDI.
Namely, an assembly in a regime that the remanence and coercivity vanish, but the field-dependent magnetization varies with concentration of MNPs indicating the presence of DDI.
The authors (Allia 2001) suggested that the dipolar field changes at a high rate and in random direction and therefore acts similar to the thermal field. The effect is accounted for by an apparent increase of the system temperature. The magnetization at temperature $T$ is given by
$M=M_s\L\left[\mu H /k_B(T+T^*) \right]$,
with $T^*$ related to the average dipolar energy via
$k_BT^*=n_1\mu^2/\overline{d^3}$ and obtained by a fitting procedure.
This model interpreted successfully the magnetization behavior of Co nanoparticles in Cu matrix and established the existence of the interacting superparamagnet regime (Allia 2001).

\subsection{Numerical Techniques}\label{ssec_MCMD}
The mean-field models have the advantage of providing analytical expressions suitable for extracting system parameters from the experimental data by a fitting process. However, they are not applicable to strongly dipolar systems and they do not account for collective effects.
Numerical techniques on the other hand, have the major advantage that they treat rigorously the local and temporal statistical fluctuations of the macroscopic quantities characterizing the MNP assembly and provide an efficient interpolation scheme between the weak and the strong interaction regimes.
We discuss briefly two most common numerical approaches, the Monte Carlo (MC) method and the Magnetization Dynamics (MD) method.
\par
\subsubsection{The Monte Carlo method}
Different algorithms that mimic thermal fluctuations of the degrees of freedom of a physical system by means of (pseudo)random numbers go under the umbrella of Monte Carlo techniques.
In the case of MNPs, two widely used algorithms are the Metropolis Monte Carlo (MMC) and the Kinetic Monte Carlo (KMC).
The former is appropriate for a description of the equilibrium behavior of an assembly, while the latter also accounts, within a certain time scale, for the transition to equilibrium.
Both algorithms provide thermal averages of macroscopic quantities of interest in the canonical ensemble, i.e. at constant temperature. To do so a sampling of the phase space is performed, however the sampling procedures differ, as outlined below.

The MMC algorithm samples the phase space, visiting preferentially states close to the equilibrium states (\textit{Importance Sampling}). This is achieved  when subsequently visited states form a Markov chain, meaning that the probability of visiting the next state depends only on the last visited one.
To do so, one chooses the transition from state $s$ to $s'$ to occur with certainty, if it reduces the total energy ($E_{s'} \le E_s$) and with a finite probability
$p(s\rightarrow s') = \exp(-\frac{E_{s'}-E_s}{kT})$,
if it increases the total energy ($E_{s'} > E_s$). Thus the system is allowed to climb-up energy barriers and slide-down toward energy minima until it reaches eventually the global minimum.

In KMC the system jumps from a state $s$ at a local minimum to a new state $s'$ being also a local minimum by overcoming a barrier $E_b$.
The jump is performed within a predefined time step $\Delta t$  with probability
$p(\Delta t)=1-\exp(-\Delta t/\tau)$
where $\tau$ is the corresponding relaxation time with Arrhenius behavior,
$\tau=\tau_0\exp(E_b/kT)$.

In both algorithms, interparticle interactions are included by replacing the applied field $H$ with the \textit{total} field
$H_i = H + \sum_{j(\ne i)} H_{int,ij}$, which includes the contribution from  the \textit{interaction} field  $H_{int,ij}$. 
In contrast to mean-field theories, in MC and MD (see next section) techniques the interaction field is treated exactly, meaning that its value depends on the configuration of all the moments of the assembly and it changes at each time-step.

An important distinction between the MC algorithms is that KMC simulates the relaxation of the system in physical time, while time quantification of the MMC time step is possible only in the absence of interparticle interactions (Nowak 2000, Chubykalo 2003).
However, a serious difficulty in KMC arises from the calculation of the local energy barrier required to obtain the transition probability.
In an interacting system the barrier depends on \textit{all} degrees of freedom and its calculation is a formidable task (Chubykalo 2004, Jensen 2006), usually performed in an approximate manner (Pfeiffer 1990, Chantrell 2001).
Furthermore, the KMC assumes that the system evolves through thermally activated jumps over energy barriers, an approximation that becomes invalid at elevated temperatures $(kT\sim E_b)$, or when collective behavior becomes important, as, for example, in strongly interacting MNPs. Collective effects are better described within the MMC algorithm.

For a detailed description and technical implementation of MC algorithms the interested reader could refer to the book by Landau and Binder (Landau 2000)
\par
\subsubsection{The Magnetization Dynamics method}
In this method, the equations of motion for the magnetic moments are integrated in time and time averages of the macroscopic magnetization are recorded.
At zero temperature, the time-evolution of a magnetic moment $\mu_i$ under a total field $H_i$ is described by the Landau-Lifshitz-Gilbert (LLG) equation
\begin{equation}
 \frac{d \overrightarrow\mu_i }{dt}=
- A ( \overrightarrow\mu \times \overrightarrow H_i )
- B_i
  \overrightarrow\mu_i \times
  ( \overrightarrow\mu_i \times \overrightarrow H_i  )
\label{eq_LLG}
\end{equation}
with
$A \equiv \gamma / (1+\alpha^2) $,
$B_i \equiv \alpha \gamma / (1+\alpha^2) \mu_i $,
$\gamma$ the gyromagnetic ratio, and
$\alpha$ a dimensionless damping parameter.
The first term on the r.h.s. of Eq.(\ref{eq_LLG}) is the torque term leading to precession around the field axis and the second one is a phenomenological damping torque that tends to align the precessing moment with the field $H_i$.

The dynamics at finite temperature are described by introduction of an additional field ($H_{f,i}$) term in Eq.(\ref{eq_LLG}) with stochastic character. $H_f$ is assumed to have zero time-average (\textit{white noise}) and its values at different sites $i,j$ or different instants $t,t'$ are uncorrelated. The LLG equation augmented by the thermal field term is commonly referred to as the \textit{Langevin} or \textit{stochastic LLG} equation.
\par
\subsubsection{Time scale of numerical methods}
The MC and MD techniques are complementary since they describe thermal relaxation of magnetic properties in different time scales.
In the MD method, the characteristic time is a fraction ($\sim10^{-2}$) of the precessional (Larmor) period $(\sim10^{-10}s)$, which implies that simulation times up to $\sim 1~ns$ are presently attainable.
Thus, MD is the appropriate scheme to investigate fast-relaxation phenomena as, for example, the reversal path of magnetization under an applied short field pulse (Berkov 2002, Suess 2002).

In KMC the characteristic time is the single-particle relaxation time (see Eq.(\ref{eq_tau})), which is much larger than $\tau_0 \sim 10^{-10}s$ in the temperature range of interest $(kT \ll E_b)$, a fact that makes the method suitable to treat slow-relaxation problems, as, for example the thermal decay of magnetization in permanent magnets, a phenomenon that evolves within days or years (Van de Veerdonk 2002).

Finally, when static magnetic properties are concerned, the system is at a stable (or metastable) state and the MMC is a powerful and sufficient scheme to describe, for example, long range order at the ground state or collective behavior at finite temperature (Kechrakos 1998, Jensen 2003).
\par
\subsubsection{MMC study of dipolar interacting assemblies : A case study}
In this section we show typical results from MMC simulations of the magnetic properties of dipolar interacting MNP assemblies (Kechrakos 1998, Kechrakos 2002).
Our system contains $N$ identical SD NPs with diameter $D$ and uniaxial anisotropy in a random direction.
The MNPs are located randomly in space or on the vertices of a hexagonal lattice.
The former is an appropriate model for granular samples, and the latter for self-assembled MNPs.
The total energy of the system is
\begin{eqnarray}
E = g\sum_{ij}\frac{ \widehat{S}_{i}  \cdot \widehat{S}_{j}
 - 3(\widehat{S}_{i} \cdot \widehat{R}_{ij} )
    (\widehat{S}_{i} \cdot \widehat{R}_{ij} )  } {R_{ij}^{3}}
 - k \sum_{i} ( \widehat{S}_{i}  \cdot \widehat{e_{i}} )^{2}
 - h \sum_{i} ( \widehat{S}_{i}  \cdot \widehat{H} )
\label{eq_Etot}
\end{eqnarray}
where $\widehat{S}_{i}$ is the magnetic moment direction (spin) of the $i$-th particle, $\widehat{e_{i}}$ is the easy-axis direction, and ${R}_{ij}$ is the center-to-center distance between particles $i$ and $j$.
Hats indicate unit vectors.
The energy parameters entering Eq.(\ref{eq_Etot}) are:
(i) the dipolar energy $g \equiv \mu_0^2\mu^{2} / 4 \pi d^{3}$, with $\mu = M_{s}V$ the particle moment and $d$ the minimum interparticle distance.
(ii) the anisotropy energy $k\equiv K_{1}V$, and
(iii) the Zeeman energy $h\equiv \mu_0\mu H$ due to the applied dc field $H$.
The energy parameters ($g,k,h$) entering Eq.(\ref{eq_Etot}), the thermal energy $k_{B}T$, and the treatment history of the sample determine the micromagnetic configuration at a certain temperature and field.
The freedom to choose an arbitrary energy scale makes the numerical results applicable to a class of materials with the same parameter ratios rather than to a specific material.
The crucial parameter that determines the transition from single-particle to collective behavior is the ratio of the dipolar to the anisotropy energy ($g/k$).
\begin{figure}
\begin{center}
\includegraphics{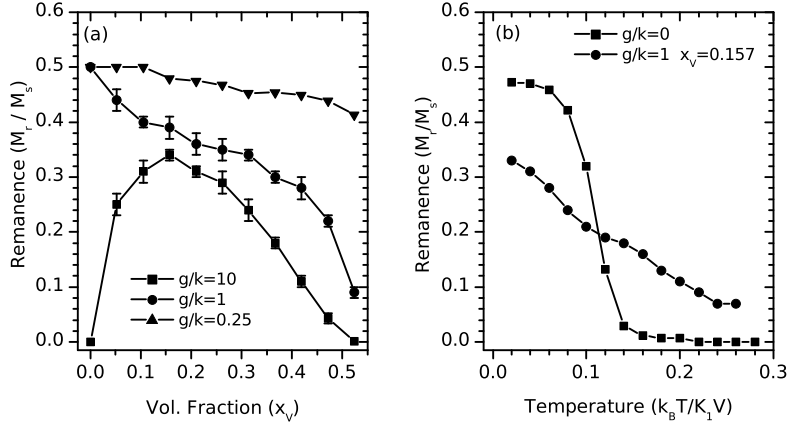}
\caption{Dependence of the saturation isothermal remanence of a random assembly (a) on volume fraction of MNPs, at very low temperature (t/k=0.001) and, (b) on temperature, for fixed volume fraction. The particles have random anisotropy. The data are obtained by MMC simulations.}
\label{fig_MrX_MrT}
\end{center}
\end{figure}
\par
We show in Fig.~\ref{fig_MrX_MrT} the concentration and temperature dependence of the remanence magnetization of a random assembly. Notice in  Fig.~\ref{fig_MrX_MrT}a, that weak DDI produce an increase of the remanence with concentration, while strong DDI have the opposite effect. Remarkably, the presence of free sample boundaries, can reverse the increasing trend of the remanence, due to the presence of a demagnetizing field. When DDI are much stronger than single-particle anisotropy ($g/k\sim10$), the remanence value is sensitive to the morphology of the assembly, as the peak around the percolation threshold indicates. This behavior is explained by the anisotropic character of DDI (see Fig.~\ref{fig_DDI}).
In Fig.~\ref{fig_MrX_MrT}b, DDI interactions are shown to produce a much slower temperature decay, producing finite remanence values above the blocking temperature of the isolated MNPs. This result supports the predictions of the mean-field theory about the increase of the measured blocking temperature in dipolar interacting systems (Dormann 1988).
\begin{figure}
\begin{center}
\includegraphics{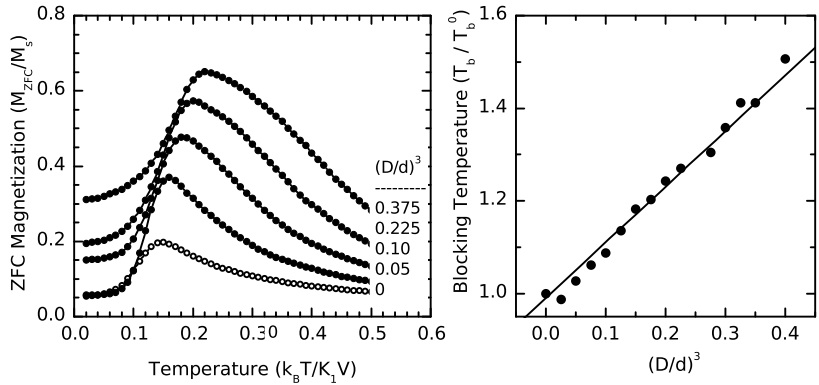}
\caption{ ZFC curves and blocking temperature for an ordered (hexagonal) assembly of identical MNPs with diameter $D$ and center-to-center distance $d$. (a) Evolution of ZFC curves with decreasing $d$ values. (b) Linear scaling of $T_b$ with inverse cube of $d$. Data obtained by MMC simulations.}
\label{fig_ZFC_Tb}
\end{center}
\end{figure}
\par
In chemically-prepared, self-assembled MNPs the possibility to control the interparticle separation by variation of the surfactant (Willard 2004) offers the possibility to study the dependence of $T_b$ on interparticle spacing while preserving the geometrical arrangement of the assembly (hexagonal).
In Fig.~\ref{fig_ZFC_Tb} we show results for the ZFC magnetization ($M_{ZFC}$) and the blocking temperature as obtained from the peak of the $M_{ZFC}(T)$ curve for a hexagonal array of dipolar interacting MNPs with random anisotropy. Parameters corresponding to Co nanoparticles are used (Kechrakos 2002).
The characteristic dependence of $T_b$ on the inverse cube of interparticle spacing can be used as a proof of the dominant character of DDI in an assembly.
Notice also that $M_{ZFC}(T\approx 0)$ assumes a positive value that increases with $d$ values. This feature arises from the gradual formation of a long range ferromagnetic ground state, due to DDI.
\par
More examples of MC or MD simulations and comparison to experiments on MNP assemblies can be found in the relevant literature (Vedmedenko 2007, Kechrakos and Trohidou 2008).

\section{Summary}  \label{sec_sum}

We have discussed the main theoretical concepts that pertain to the magnetization properties of isolated (non-interacting) nanoparticles and their assemblies. We estimated the critical radius for formation of single domain particles and studied the zero-temperature magnetization reversal mechanism of coherent rotation (Stoner-Wohlfarth model), the thermally activated reversal (N\'{e}el-Brown model) and the related phenomenon of superparamagnetism occurring above the blocking temperature.
Complications arising from size polydispersity, distribution of easy axes directions, and applied field on the relaxation time for magnetization reversal were discussed.
Two standard experimental techniques for (static) magnetic measurements, namely the field and temperature dependence of magnetization were outlined. Finally, the subject of interparticle dipolar interactions was introduced along with the most common theoretical techniques used to analyze interacting systems. Examples from Monte Carlo studies of MNP assemblies were given.

\section{Future perspectives}  \label{sec_persp}

The dynamic behavior of MNPs in the presence of interparticle interactions is expected to remain a topic of intense scientific and technological research in the coming years.
The research effort is expected to focus on both the atomic scale properties of individual magnetic nanoparticles and on the mesoscopic properties of nanoparticle assemblies.

On the \emph{atomic} scale, future goals will include~: 
(i) Reduction of the magnetic particle size without violating thermal stability (\emph{superparamagnetic limit}) at room temperature . The technological benefit from progress in this direction will be the development of magnetic data-storage media with higher areal density. Given that the SPM effect is not observed below a certain size of a MNP, due to disorder effects on the particle surface, the search for new high-anisotropy materials is required. Composite nanoparticles with a core-shell morphology (Skumryev 2003) constitute an interesting perspective.\\
(ii) Understanding and control of surface effects. With reduction of particle size the contribution from surface moments become of increasing importance. The chemical structure of the surface (disorder, defects) controls the magnitude and type of the surface anisotropy, which is usually much  (up to $\sim 10$ times) larger than the core anisotropy. Synthetic methods can offer indispensable routes to surface structure modification. \emph{Ab-initio} electronic structure calculations are a valuable tool to predict the surface anisotropy values and modeling of MNPs as multi-spin system will reveal complex magnetization reversal mechanisms beyond the Stoner-Wohlfarth model
(Kachkachi 2000). Experiments on individual nanoparticles (Wernsdorfer 2000)
offer a unique test of the above theories.

The future task on the \emph{mesoscopic} scale will be to
understand and control collective magnetic behavior in ordered nanostructures (self-assembled MNPs and magnetic patterned media).
Ordered nanostructures include chemically prepared self-assembled MNPs and lithographically prepared magnetic patterned media.  
Chemical synthesis of MNPs and self-assembly (bottom-up approach) is a very promising and cost-effective method to produce ordered MNP arrays (Willard 2004). 
However, deeper understanding and improvement of the self-assembly  process is required in order to achieve larger (beyond $1mm^2$) sample area with structural coherence. 
There is still a remaining problem as nanoparticles self-assemble into hexagonal arrays that are incompatible with the square arrangements required in industrial applications. 
A resolution to this problem could be the recently demonstrated templated assembly (Cheng 2004).
Lithographic patterning (top-down approach) offers better control over the geometrical aspects of the assembly but cannot yet produce nanostructures with size below $\sim 100nm$ (Martin 2003).
Increase of lithographic resolution is demanded in order to achieve patterned media with smaller (below $0.1\mu m$) characteristic size.
On the measurements side, improvement of existing techniques to probe mesoscopic magnetic order and excitations is demanded. Recent examples are the observation of mesoscopic sale magnetic order in self-assembled Co nanoparticles by an indirect method (small-angle neutron scattering)(Sachan 2008) and by direct methods such as magnetic force microscopy (Puntes 2004)  and electron holography (Yamamoto 2008).
From the point of view of basic physics, ordered nanostructures constitute model systems to study collective magnetic behavior driven by magnetostatic interactions, because the size, the shape and the spatial arrangement of the magnetic nanostructures is well controlled. Known phenomena are to be demonstrated on the mesoscopic scale and new ones possibly to be discovered. As a recent example, we refer to the observation of magnetic frustration is magnetostatically coupled magnetic microrods (Wang 2006), a phenomenon previously met in bulk magnetic random alloys (spin glasses).
\par
Finally, progress in numerical modeling will provide methods for bridging the atomic scale and the  mesoscopic scale simulations. Such multi-scale simulations point to the future of theoretical investigations in the field of relaxation in magnetic nanoparticles and have only recently started to appear (Yanes 2007, Kazantseva 2008).

\section*{References}

\begin{description}

    \item Aharoni, A. 1965.
Effect of a Magnetic Field on the Superparamagnetic Relaxation Time.
\textit{Phys. Rev.} 177: 793.

    \item Allia, P., Coisson, M., Tiberto, P., \textit{et al}. 2001.
Granular Cu-Co alloys as interacting superparamagnets.
\textit{Phys. Rev. B} 64: Art. No. 144420

    \item Bean, C. P. and Jacobs, I. S. 1956.
Magnetic Granulometry and Super-Paramagnetism.
\textit{J. Appl. Phys.} 27: 1448-1452.

    \item Bean, C. P. and Livingston, J. D. 1959.
Superparamagnetism.
\textit{J. Appl. Phys.} 30: 120S-129S.

    \item Berkov, D.V. 2002.
Fast switching of magnetic nanoparticles: Simulation of thermal noise effects using the Langevin dynamics
\textit{IEEE Trans. Magn.} 38(5): 2489-2495

    \item Brown, W. F. Jr. 1963.
Thermal fluctuations of single-domain particle.
\textit{Phys. Rev.} 130: 1677-1686.

    \item Cheng J. Y., Mayes, A. M. and Ross, C. A. 2004.
Nanostructure engineering by templated self-assembly of block copolymers.
\textit{Nat. Mater.} 3: 823-828.

    \item Chantrell, R. W., Walmsley, N., Gore J., \textit{et al}. 2001
Calculations of the susceptibility of interacting superparamagnetic particles
\textit{Phys. Rev. B} 63 (2): Art. No. 024410

    \item Chubykalo, O., Nowak, U., Smirnov-Rueda, R., \textit{et al}. 2003.
Monte Carlo technique with a quantified time step : Application to the motion of magnetic moments.
\textit{Phys. Rev. B} 67: Art. No. 064422

    \item Chubykalo-Fesenko, O. A. and Chantrell, R. W. 2004.
Numerical evaluation of energy barriers and magnetic relaxation in interacting nanostructured magnetic systems.
\textit{J. Magn. Magn. Mater.} 343: 189-194

    \item Cowburn, R. P. 2006.
Where Have All the Transistors Gone ?
\textit{Science} 311 : 183-184.

    \item Cullity, B. D. 1972.
\textit{Introduction to Magnetic Materials}.
Addison-Wesley Publishing Company.

    \item Coffey, W. T., Crothers, D. S. F. , Dormann, J. L., \textit{et al}. 1998.
Effect of an oblique magnetic field on the superparamagnetic relaxation time : Influence of the gyromagnetic term.
\textit{Phys. Rev. B} 58: 3249-3266.

    \item Coffey, W. T., Cregg, P. J. and Kalmykov, Yu. P. 1993.
On the Theory of Debye and N\'{e}el Relaxation of Single Domain Ferromagnetic Particles.
\textit{Adv. Chem. Phys.} 69: 263-315.

    \item Darling, S. B. and Bader, S. D. 2005.
A materials chemistry perspective on nanomagnetism.
\textit{J. Mater. Chem.} 15: 4189-4195.

    \item Dormann, J. L., Bessais, L. and Fiorani, D. 1988.
A dynamic study of small interacting particles: superparamagnetic model and spin-glass laws.
\textit{J. Phys. C: Solid State Phys.} 21: 2015-2034.

    \item Dormann, J. L., Fiorani, D., and Tronc, E. 1997.
Magnetic relaxation in fine-particle systems.
\textit{Adv. Chem. Phys.} 98: 283-494.

    \item Farrell, D., Cheng, Y., McCallum, R. W., \textit{et al}. 2005.
Magnetic interactions of iron nanoparticles in arrays and dilute dispersions.
\textit{J. Phys. Chem. B} 109: 13409-13419.

    \item Garanin, A., Kennedy, E. C., Crothers, D. S. F., \textit{et al}. 1999.
Thermally activated escape rates of uniaxial spin systems with transverse field: Uniaxial crossovers.
\textit{Phys. Rev. E} 60: 6499.

    \item Hansen, M. F. and M{\o}rup, S. 1988.
Models for the dynamics of interacting magnetic nanoparticles.
\textit{J. Magn. Magn. Mater.} 184: 262-274.

    \item Held, G. A., Grinstein, G., Doyle, H., \emph{et al}. 2001.
Competing interactions in dispersions of superparamagnetic nanoparticles. \textit{Phys. Rev. B} 64: Art. No. 012408.

    \item Jensen, P. J. and  Pastor, G. M. 2003.
Low-energy properties of two-dimensional magnetic nanostructures: interparticle interactions and disorder effects.
\textit{New J. Phys.} 5: Art. No. 68

    \item Jensen, P. J. 2006.
Average energy barriers in disordered interacting magnetic nanoparticles.
\textit{Comp. Mater. Science} 35: 288-291

    \item Kazantseva, N., Hinzke, D., Nowak, U., \textit{et al}. 2008.
Towards multiscale modeling of magnetic materials: Simulations of FePt.
\textit{Phys. Rev. B} 77: Art. No. 184428.

	\item Kachkachi H, Ezzir A, Nogues M, \emph{et al}. 2000.
Surface effects in nanoparticles: application to maghemite $\gamma$-Fe$_2$O$_3$
\textit{Eur. J. Phys. B} 14(4): 681-689.

    \item Kechrakos, D. and Trohidou, K. N. 1998.
Magnetic properties of dipolar interacting single-domain particles.
\textit{Phys. Rev. B} 58: 12169-12177.

    \item Kechrakos, D. and Trohidou, K. N. 2002.
Magnetic properties of self-assembled interacting nanoparticles.
\textit{Appl. Phys. Lett.} 81: 4574-4576.

    \item Kechrakos, D., and Trohidou, K.N. 2008. 
Dipolar interaction effects in the magnetic and magnetotransport properties of ordered nanoparticle arrays. 
\textit{J. Nanoscience Nanotech.} 8(6): 2929-2943.

    \item Kodama, R. H. 1999.
Magnetic nanoparticles.
\textit{J. Magn. Magn. Mater.} 200: 359-372.
    
    \item Knobel, M., Nunes, W. C., and  Sokolovsky, L. M., \textit{et al}. 2008.
Superparamagnetism and other magnetic features in granular materials.
\textit{J. Nanosci. Nanotechnol.} 8: 2836-2857

    \item Landau, D. P. and Binder, K. 2000.
\textit{A Guide to Monte Carlo Simulations in Statistical Physics.}
Cambridge University Press.

    \item Mart\'{\i}n, J. I., Nogu\'{e}s, J, Liu, K., et al. 2003.
Ordered magnetic nanostructures: Fabrication and properties.
\textit{J. Magn. Magn. Mater.} 256: 449-501

    \item Moser, A., Takano, K., Margulies, D. T., \textit{et al}. 2002.
Magnetic recording: advancing into the future.
\textit{J. Phys. D.: Appl. Phys.} 35: R157-R167.

    \item Murray, C. B., Sun, S., Doyle, H., \emph{et al}. 2001.
Monodisperse 3d transition-metal (Co, Ni, Fe) nanoparticles and their assembly into nanoparticle superlattices.
\textit{MRS Bull.} 26: 985-991.

    \item N\'{e}el, L. 1949.
Influence de fluctuations thermiques sur l' aimantation de grains ferromagn\'{e}tiques tr\'{e}s fins.
\textit{Compt. Rend.} 228: 664-666

    \item Nowak, U., Chantrell, R. W., and Kennedy, E. C. 2000.
Monte Carlo simulation with time-step quantification in terms of Langevin dynamics.
\textit{Phys. Rev. Lett.} 84: 163-166.

    \item Nunes, W. C., Folly, W. S. D. , Sinnecker, J. P., and Novak, M. A. 2004. 
Temperature dependence of the coercive field in single-domain particle systems. 
\textit{Phys. Rev. B }70: Art. No. 014419.

    \item O'Grady, K.,  El-Hilo, M., and Chantrell, R.W. 1993. The characterization of interaction effects in fine-particle systems. 
    \textit{IEEE Trans. Magn.} 29(6): 2608-2613.

    \item Petit, V., Taleb, A., and Pileni, M.P. 1998.
Self-organization of magnetic nanosized cobalt particles.
\textit{Adv. Mater.} 10: 259-261.

    \item Pfeiffer, H. 1990.
Determination of anisotropy field distribution in particle assemblies taking into account thermal fluctuations.
\textit{Phys. Stat. Sol. (a)} 118: 295-306.

    \item Puntes, V. F., Krishnam, K. M., and Alivisatos, A.P. 2001.
Colloidal nanocrystal shape and size control: The case of cobalt.
\textit{Science} 291: 2115-2117.

    \item Puntes, V. F., Gorostiza, P., Aruguete, D. M., \emph{et al}. 2004.
Collective behavior in two-dimensional cobalt nanoparticle assemblies observed by magnetic force microscopy.
textit{Nat. Mater.} 3: 263-268.

    \item Russier, V., Petit, C., Legrand, J., \emph{et al}. 2000.
Collective magnetic properties of cobalt nanocrystals self-assembled in a hexagonal network: Theoretical model supported by experiments.
\textit{Phys. Rev. B} 62: 3910-3916.

    \item Russier, V., Petit, V. and Pileni, M.P. 2003.
Hysteresis curve of magnetic nanocrystals monolayers: Influence of the structure.
\textit{J. Appl. Phys.} 93: 10001-10010.

    \item  Sachan M., Bonnoit, C., Majetich, S. A., \emph{et al}. 2008.
Field evolution of magnetic correlation lengths in $\epsilon$-Co nanoparticle assemblies.
\textit{Appl. Phys. Lett.} 92(15): Art. No. 152503.

    \item Sasaki, M., J\"{o}nsson, P.E., Takayama, H., \emph{et al}. 2005.
Aging and memory effects in superparamagnets and superspin glasses.
\textit{Phys. Rev. B} 71(10): Art. No. 104405.

    \item Skumryev, V., Stoyanov, S., Zhang, Y. , \emph{et al}. 2003.
Beating the superparamagnetic limit with exchange bias.
\textit{Nature} 423: 850-853.

    \item Stoner, E. C. and Wohlfarth, E. P. 1948.
A mechanism of magnetic hysteresis in heterogeneous alloys.
\textit{Proc. R. Soc. London A} 240: 599-642.

    \item Shtrikman, S. and Wohlfarth, E. P. 1981.
The theory of Vogel-Fulcher law of spin glasses.
\textit{Phys. Lett. A} 85(8,9): 467-470.

    \item Suess, D., Schrefl, T., Scholz, W., \textit{et al}. 2002
Fast switching of small magnetic particles
\textit{J. Magn. Magn. Mater.} 242: 426-429.

    \item Van de Veerdonk, R. J. M., Wu, X. W., Chantrell, R. W., \textit{et al}. 2002.
Slow dynamics in perpendicular media.
\textit{IEEE Trans. Magn.} 38(4): 1676-1681

    \item Vedmedenko, E. Y. 2007.
\textit{Competing Interactions and Patterns in Nanoworld}.
Wiley-VCH Verlag.

    \item Wang, R.F., Nisoli, C., Freitas, R.S., \textit{et al}. 2006.
Artificial 'spin ice' in a geometrically frustrated lattice of nanoscale ferromagnetic islands.
\textit{Nature} 439(7074): 303-306.

    \item Wernsdorfer, W., Mailly, D. and Benoit, A. 2000.
Single nanoparticle measurement techniques.
textit{J. Appl. Phys.} 87(9): 5094-5096.

    \item Willard, M. A., Kurihara, L. K., Carpenter, E. E., \textit{et al}. 2004.
Chemically prepared magnetic nanoparticles
\textit{Int. Mater. Rev.} 49 : 125-170

    \item Wohlfarth, E. P. 1958.
Relations between different modes of acquisition of the remanent magnetization of ferromagnetic particles.
\textit{J. Appl. Phys.} 29: 595-596

    \item Woods, S. I., Kirtley, J. R., Sun, S., \emph{et al}. 2001.
Direct investigation of superparamagnetism in Co nanoparticle films. \textit{Phys. Rev. Lett.} 87: 137205-137208.

    \item Yamamoto, K., Majetich, S.A., McCartney, M.R., \textit{et al}. 2008.
Direct visualization of dipolar ferromagnetic domain structures in Co nanoparticle monolayers by electron holography.
\textit{Appl. Phys. Lett.} 93(8): Art. No. 082502

    \item Yanes, R., Chubykalo-Fesenko, O., Kachkachi, H., \textit{et al}. 2007.
Effective anisotropies and energy barriers of magnetic nanoparticles with N\'{e}el surface anisotropy.
\textit{Phys. Rev. B} 76(6): Art. No. 064416

\end{description}


\end{document}